\documentclass[12pt]{iopart}
\usepackage[dvips]{graphics}
\usepackage{graphicx}
\usepackage{amsfonts}
\usepackage{amssymb}
\usepackage{subfigure}
\usepackage{cite}
\usepackage{color}

\def\rt{({\bf r},t)}
\def\rtau{({\bf r},t-\tau)}

\def\ro{({\bf r},\omega)}

\def\zt{(z,t)}

\begin{document}

\title[Vector solitons in nonlinear isotropic chiral metamaterials]
{Vector solitons in nonlinear isotropic chiral metamaterials}

\author{N.~L.~Tsitsas$^1$, A.~Lakhtakia$^2$, and D.~J.~Frantzeskakis$^3$}

\address{$^1$
School of Applied Mathematical and Physical Sciences,
National Technical University of Athens, Zografos, Athens 15773, Greece}

\address{$^2$
Department of Engineering Science and Mechanics, Pennsylvania State University, University Park, PA 16802-6812, USA}

\address{$^3$
Department of Physics, University of Athens, Panepistimiopolis, Zografos, Athens 15784, Greece }

\begin{abstract}
Starting from the Maxwell equations, we used the reductive perturbation method to derive a system of two coupled nonlinear Schr\"{o}dinger (NLS) equations
for the two Beltrami components of the electromagnetic field propagating along a fixed direction in an isotropic nonlinear chiral metamaterial.
With single-resonance Lorentz models for the permittivity and permeability and a Condon model for the chirality parameter, in certain spectral regimes,
one of the two Beltrami components exhibits a negative-real refractive index when nonlinearity is ignored and the chirality
parameter is sufficiently large. We found that, inside such a spectral regime, there may exist a subregime wherein the system of the NLS equations
can be approximated by the Manakov system. Bright-bright, dark-dark and dark-bright vector solitons can be formed in that spectral subregime.

\end{abstract}

\maketitle

\section{Introduction}

Metamaterials exhibiting a refractive index with a negative real
part in certain spectral regimes have been a topic of intense
research activity over the last decade
\cite{Review1,Review2,Review3}. A sufficient condition for the
exhibition of a negative-real refractive index (NRRI) by an
isotropic dielectric-magnetic material is that both its
permittivity $\tilde{\epsilon}(\omega)$ and permeability
$\tilde{\mu}(\omega)$ have negative real parts for the same value
of the angular frequency $\omega$ \cite{DL2004}. That demanding
condition can be relaxed if the isotropic material possesses
chirality. In this case, one of the two refractive indices
exhibited by such an isotropic chiral material can have a negative
real part \cite{ml2004,pendrych};
besides, the possibility of realizing NRRI materials through the
routes of gyrotropy and nonreciprocity \cite[Sec.~2.3.4]{MLbook}
was considered in the more general context of Faraday chiral
materials \cite{ml2004}.
Although the electromagnetic (EM) properties of isotropic chiral
materials had been studied in detail in the recent past
\cite{bookchiral,bookchiral1994,bookchiral2}, the possibility of
isotropic chiral NRRI materials gave fresh impetus \cite{lakrev}.
Significant theoretical progress has been reported
\cite{ICMtheo1,ICMtheo2,ICMtheo3,ICMtheo4,ICMtheo5} in comparison
to experimental progress \cite{ICMexp1}.

All of the foregoing NRRI materials have been considered to have linear response properties.
Nonlinear NRRI materials---of the isotropic achiral type---were introduced \cite{ZSK2003,ASBZ2004} shortly after their linear counterparts. Theoretical
research has been extensively performed on these nonlinear NRRI materials, as they feature nonlinearity-induced localization of
EM waves and soliton formation \cite{nlmms1a,nlmms1b,nlmms1c,nlmms1d,nlmms1e,nlmms2a,nlmms2b,nlmms2c}.
Similar effects had been predicted earlier for nonlinear isotropic chiral materials with positive-real refractive
indices \cite{BSML1995,SMBL1995,HK1995,FSY2002}.
The possible incorporation of nonlinearity in isotropic chiral NRRI materials motivated us to investigate soliton formation in these materials.

In this paper, we report on the propagation of an EM field along a
fixed direction in an isotropic chiral NRRI material with
nonlinear permittivity and permeability of the Kerr type
\cite{Weinberger}. Furthermore, our analytical approximations may
also be useful for analyzing field propagation along the
distinguished axis of a nonlinear uniaxial bianisotropic material.
We start from the Maxwell equations in the
time domain and use the reductive perturbation method (RPM)
\cite{rpm1,rpm2} to derive a system of two coupled nonlinear
Schr\"{o}dinger (NLS) equations for the left-handed and
right-handed Beltrami components of the EM field
\cite{bookchiral1994}. Then, we adopt (i) the Lorentz model for
the linear parts of the relative permittivity and permeability and
(ii) the Condon model for the (linear) chirality parameter; see,
e.g., Ref.~\cite{lakhtakia-josaa} and references therein.

Now, if the chirality parameter is sufficiently large in a certain spectral regime and
both dissipation and nonlinearity are ignored, the refractive index for the left/right-handed Beltrami
component is real and negative but that for right/left-handed Beltrami component is real and positive \cite{ICMtheo2}. Inside that spectral regime, perturbative nonlinearity of the chosen kind induces a spectral subregime  wherein
the system of the NLS equations can be approximated by the Manakov system \cite{manakov}. Since the latter is known to be completely integrable \cite{intman1,intman2,intman3}, we can predict various classes of exact
vector soliton solutions that can be supported in the nonlinear isotropic chiral NRRI material: depending on the type of nonlinearity, namely self-focusing or self-defocusing, we find that bright-bright solitons (for the focusing case), as well as dark-dark and dark-bright solitons (for the defocusing case) can be formed. In all cases, these vector solitons are composed of a Beltrami component with negative-real refractive index and another with a positive-real refractive index.

The paper is organized as follows. In Sec.~\ref{sec2} we present the constitutive relations and use the RPM to derive, from
the time-domain Maxwell equations, the system of two coupled NLS equations. In Sec.~\ref{sec3} we show that the coupled NLS
equations can be approximated by the Manakov system, and we present various types of vector soliton solutions. We conclude in Sec.~\ref{sec4}.

\section{Coupled nonlinear Schr\"odinger equations}\label{sec2}

\subsection{Constitutive relations}\label{conrels}

In the time domain, the Tellegen constitutive relations of an isotropic chiral material are expressed  as \cite{bookchiral,bookchiral1994,bookchiral2}
\begin{eqnarray}
&&
\mathbf{D}\rt= \int_{-\infty}^{t}\left[
\epsilon(\tau)\ast\mathbf{E}\rtau +\xi(\tau)\ast\mathbf{H}\rtau\right]d\tau \,,
\label{eq:constitutive-relations-1}
\\[5pt]
&&
\mathbf{B}\rt= \int_{-\infty}^{t}\left[\mu(\tau)\ast\mathbf{H}\rtau
-\xi(\tau)\ast\mathbf{E}\rtau\right]d\tau\,,
\label{eq:constitutive-relations-2}
\end{eqnarray}
where $\mathbf{D}\rt$ and $\mathbf{B}\rt$ are the electric flux density
and the magnetic induction field, respectively; $\epsilon(t)$ is the inverse Fourier
transform of $\tilde{\epsilon}(\omega)$, and $\mu(t)$
of $\tilde{\mu}(\omega)$; and
$\xi(t)$, the inverse Fourier transform of $\tilde{\xi}(\omega)$,  accounts for the chirality of the chosen material.
Hereafter, ${\bf r}=x\hat{\bf {x}}+y\hat{\bf {y}}+z\hat{\bf {z}}$ is the position vector and $t$ denotes time.

We specify the permittivity and the permeability in the frequency domain in terms of linear and nonlinear
components as follows:
\begin{eqnarray}
\label{eq:e-m-1}
\tilde{\epsilon}(\omega)&=&\tilde{\epsilon}_\ell(\omega)+\tilde{\epsilon}_{n\ell}(|\mathbf{E}|^2,|\mathbf{H}|^2),\quad
\quad
\\
\tilde{\mu}(\omega)&=&\tilde{\mu}_\ell(\omega)+\tilde{\mu}_{n\ell}(|\mathbf{E}|^2,|\mathbf{H}|^2),\qquad
\label{eq:e-m-2}
\end{eqnarray}
The linear components of $\tilde{\epsilon}_\ell$ and $\tilde{\mu}_\ell$ depend on $\omega$ and can be decomposed into real
and and imaginary parts, i.e.,
\begin{eqnarray}
\tilde{\epsilon}_\ell(\omega) &=& \tilde{\epsilon}_{R}(\omega)+i\tilde{\epsilon}_{I}(\omega),
\label{reim1} \\
\tilde{\mu}_\ell(\omega) &=& \tilde{\mu}_{R}(\omega)+i\tilde{\mu}_{I}(\omega),
\label{reim2}
\end{eqnarray}
where the imaginary parts $\tilde{\epsilon}_{I}$ and
$\tilde{\mu}_{I}(\omega)$ account for dissipation. The nonlinear
parts $\tilde{\epsilon}_{n\ell}$ and $\tilde{\mu}_{n\ell}$ are
assumed to be independent of frequency and depend quadratically
but weakly on the magnitudes of both the electric field and the
magnetic field. In particular, we consider an isotropic Kerr
nonlinearity with
\begin{eqnarray}
\tilde{\epsilon}_{n\ell} &=& \alpha_1 \vert\mathbf{E}\vert^2 + \beta_1 \vert\mathbf{H}\vert^2,
\label{enl} \\
\tilde{\mu}_{n\ell} &=& \alpha_2 \vert\mathbf{E}\vert^2 + \beta_2 \vert\mathbf{H}\vert^2,
\label{mnl}
\end{eqnarray}
where $\alpha_{1,2}$ and $\beta_{1,2}$ are scalar Kerr
coefficients governing the magnitude of the nonlinearity. These
four coefficients are the necessary ones to describe nonlinear
wave propagation, in the framework of the NLS model, in an
isotropic chiral material; see Sec.~2.3 later in this paper along
with Ref.~\cite{SMBL1995} for earlier work. The assumed form of
the nonlinear dependence of $\tilde{\epsilon}_{n\ell}$ and
$\tilde{\mu}_{n\ell}$ may stem from a Taylor expansion of a more
general type of nonlinearity (e.g., a saturable one) \cite{kiag}.

The frequency-domain chirality parameter $\tilde{\xi}(\omega)$ is
purely linear, and is defined as
\begin{eqnarray}
\tilde{\xi}(\omega) = \frac{i}{c}\tilde{\kappa}(\omega)=\frac{i}{c}
\left[\tilde{\kappa}_R(\omega)+i\tilde{\kappa}_I(\omega)\right]\,,
\label{kappa}
\end{eqnarray}
where $c$ is the speed of light in free space (vacuum), and the
frequency-dependent function $\tilde{\kappa}$ defines the chiral
properties of the material, with its imaginary part
$\tilde{\kappa}_I$ accounting for dissipation together with
$\tilde{\epsilon}_I$ and $\tilde{\mu}_I$.

Finally, we define the complex wave numbers
\begin{eqnarray}
\tilde{k}^\pm(\omega)&=& \frac{\omega}{c}\left( \sqrt{\frac{\tilde{\epsilon}(\omega)\tilde{\mu}(\omega)}
{\epsilon_0\mu_0}}\pm \tilde{\kappa}(\omega) \right) = \tilde{k}_R^\pm(\omega)+i\tilde{k}_I^\pm(\omega),
\label{dr}
\end{eqnarray}
and the complex refractive indices
\begin{eqnarray}
\tilde{n}^\pm(\omega)= \frac{c}{\omega}\tilde{k}^\pm(\omega)=
 \tilde{n}_R^\pm(\omega)+i\tilde{n}_I^\pm(\omega)\,,
\label{cri}
\end{eqnarray}
where the superscript $+$ identifies the left-handed Beltrami component and the
superscript $-$ identified the right-handed Beltrami component, whereas $\epsilon_0$
and $\mu_0$ are the permittivity and permeability of free space.
With an $\exp(-i\omega t)$ time dependence,
if $\tilde{\bf E}\ro$ and $\tilde{\bf H}\ro$
are the electric and magnetic field phasors, then the two Beltrami components given by
\begin{eqnarray}
\tilde{\bf Q}^{\pm}\ro=\frac{1}{2}\left[\tilde{\bf E}\ro {\pm} i\tilde{\eta}_\ell(\omega)\tilde{\bf H}\ro\right],
\label{belt}
\end{eqnarray}
obey the equations
\begin{eqnarray}
\nabla\times\tilde{\bf Q}^\pm\ro= {\pm}\tilde{k}^\pm(\omega)\,\tilde{\bf Q}^\pm\ro,
\label{belt2}
\end{eqnarray}
where $\tilde{\eta}_\ell^2(\omega)=\tilde{\mu}_\ell(\omega)/\tilde{\epsilon}_\ell(\omega)$, and the
nonlinear properties have been ignored \cite{bookchiral1994,bookchiral2}.

\subsection{Propagation along the $+z$ axis}
Faraday and Amp\'{e}re--Maxwell equations are, respectively, stated as
$\nabla\times{\bf E}\rt = -(\partial/\partial t){\bf B}\rt$ and $\nabla\times{\bf H}\rt = (\partial/\partial t){\bf D}\rt$.
Let us assume that the
EM field is propagating along the $+z$ axis. It then has
to be transversely polarized:
$\mathbf{E}(z,t)=\hat{\mathbf{x}} E_x(z,t)+\hat{\mathbf{y}} E_y(z,t)$ and
$\mathbf{H}(z,t)=\hat{\mathbf{x}} H_x(z,t)+\hat{\mathbf{y}} H_y(z,t)$.
Furthermore, the field has a carrier angular frequency $\omega_c$.

Following the Bohren decomposition of the
EM field phasors into left- and right-handed
Beltrami components \cite{bookchiral1994,SMBL1995} provided in Sec.~\ref{conrels}, we represent
\begin{eqnarray}
\mathbf{E}(z,t)&=\left[\hat{\mathbf{e}}_{+} q^{+}(z,t){\rm e}^{ik^+_c z}
+\hat{\mathbf{e}}_{-} q^{-}(z,t){\rm e}^{ik^-_c z}\right]{\rm e}^{-i\omega_c t}+{\rm c.c.},
\label{eq:Efield-decomp} \\
\mathbf{H}(z,t)&=\left[\hat{\mathbf{e}}_{+} p^{+}(z,t){\rm e}^{ik^+_c z}
+\hat{\mathbf{e}}_{-} p^{-}(z,t){\rm e}^{ik^-_c z}\right]{\rm e}^{-i\omega_c t}+{\rm c.c.},
\label{eq:Hfield-decomp}
\end{eqnarray}
where ``c.c.'' denotes the complex conjugate,
$k^\pm_c=\tilde{k}^\pm(\omega_c)$, $\hat{\mathbf{e}}_{\pm}=
(\hat{\mathbf{x}}\pm i\hat{\mathbf{y}})/\sqrt{2}$, whereas the
field envelopes $q^{\pm}\zt$ and $p^{\pm}\zt$ have to be
determined. We also define
$k_{c}^{\pm(1)}={d\tilde{k}^{\pm}(\omega)}/{d\omega}\vert_{\omega=\omega_c}$
and
$k_{c}^{\pm(2)}={d^2\tilde{k}^{\pm}(\omega)}/{d\omega^2}\vert_{\omega=\omega_c}$.

\subsection{Nonlinear evolution equations}
Nonlinear evolution equations for the unknown field envelopes can
be found by the reductive perturbation method \cite{rpm1,rpm2} as
follows. We introduce the slow variables
\begin{eqnarray}
\label{eq:slow-variables} Z=\varepsilon^2z, \quad \quad
T^{\pm}=\varepsilon\left(t- k_{c}^{\pm(1)}z\right),
\end{eqnarray}
where $\varepsilon$ is a formal small parameter, defined as the
relevant temporal spectral width of the nonlinear term with
respect to the spectral width of the quasi-plane-wave dispersion
relation \cite{hasbook1,hasbook2,hasbook3}. Furthermore, we
express $q^{\pm}\zt$ and $p^{\pm}\zt$ as asymptotic expansions in
terms of the parameter $\varepsilon$ as follows:
\begin{eqnarray}
q^{\pm}(Z,T^{\pm})&=q^{\pm}_0(Z,T^{\pm})+\varepsilon
q^{\pm}_1(Z,T^{\pm})+ \varepsilon^2 q^{\pm}_2(Z,T^{\pm})+ \ldots\,,
\label{eq:q-asympt}
\\
p^{\pm}(Z,T^{\pm})&=p^{\pm}_0(Z,T^{\pm})+\varepsilon
p^{\pm}_1(Z,T^{\pm})+\varepsilon^2 p^{\pm}_2(Z,T^{\pm})
+\ldots\,.
\label{eq:p-asympt}
\end{eqnarray}

Next, we substitute Eqs.~(\ref{eq:q-asympt}) and (\ref{eq:p-asympt}) as well as the constitutive relations
into the Faraday and the Amp\'{e}re--Maxwell equations, and expand $\tilde{\epsilon}_\ell(\omega)$,
$\tilde{\mu}_\ell(\omega)$, and $\tilde{\kappa}_\ell(\omega)$ about the angular frequency $\omega_c$.
 Assuming that all members of the set  $\left\{\alpha_{1,2},\beta_{1,2},\tilde{\epsilon}_{I},\tilde{\mu}_{I},\tilde{\kappa}_{I}\right\}$ are
of order $\mathcal{O}(\varepsilon^2)$, we obtain the following equations at various orders of $\varepsilon$:
\begin{eqnarray}
\mathcal{O}(\varepsilon^0): \,\,
\mathbf{W^{\pm}}\mathbf{f}^{\pm}_0&=&\mathbf{0},
\label{eq:zero-order} \\
\mathcal{O}(\varepsilon^1): \,\,
\mathbf{W^{\pm}}\mathbf{f}^{\pm}_1&=&-i(\mathbf{W^{\pm}})'\partial_T^{\pm}\mathbf{f}^{\pm}_0,
\label{eq:first-order} \\
\nonumber \mathcal{O}(\varepsilon^2):
\,\,\mathbf{W^{\pm}}\mathbf{f}^{\pm}_2&=&-i(\mathbf{W^{\pm}})'\partial_{T^{\pm}}\mathbf{f}^{\pm}_1
+\frac{1}{2}(\mathbf{W^{\pm}})''\partial^2_{T^{\pm}}\mathbf{f}^{\pm}_0
\\
&{\pm}&\frac{1}{2}k^{\pm(2)}_c
\mathbf{J}\partial^2_{T^{\pm}}\mathbf{f}^{\pm}_0\,{\mp}\,i\mathbf{J}\partial_Z\mathbf{f}^{\pm}_0
-\mathbf{A}^{\pm}_0 -\mathbf{B}\mathbf{f}^{\pm}_0.
\label{eq:second-order}
\end{eqnarray}
Here and hereafter, the prime denotes the derivative with respect
to $\omega$,
%
\begin{eqnarray}
\mathbf{f}^{\pm}_j= \left[\begin{array}{c} q^{\pm}_j\\[5pt] p^{\pm}_j\end{array}
\right]\,,\quad j\in\left\{0,1,2\right\}\,,
\end{eqnarray}
\begin{eqnarray}
\label{eq:matrix-W}
&\mathbf{W^{\pm}} = \left[ \begin{array}{cc}
\mp k^{\pm}_c+\frac{\omega_c}{c}\tilde{\kappa}_R & \frac{i\omega_c}{c}\tilde{\mu}_R\\[5pt]
\frac{i\omega_c}{c}\tilde{\epsilon}_R & \pm k^{\pm}_c-\frac{\omega_c}{c}\tilde{\kappa}_R\\
\end{array} \right],
\\[4pt]
\label{eq:matrix-A}
&\mathbf{A}^{\pm}_0 =
i\omega_c \left[
\begin{array}{c}
\mu_0[\alpha_2(|q^{+}_0|^2+|q^{-}_0|^2)+\beta_2(|p^{+}_0|^2+|p^{-}_0|^2)]p^{\pm}_0\\[5pt]
\epsilon_0[\alpha_1(|q^{+}_0|^2+|q^{-}_0|^2)+\beta_1(|p^{+}_0|^2+|p^{-}_0|^2)]q^{\pm}_0\\
\end{array} \right],
\\[4pt]
\label{eq:matrix-B}
&\mathbf{B} = -\frac{\omega_c}{c}\left[
\begin{array}{cc}
-i\tilde{\kappa}_I & \tilde{\mu}_I\\
\tilde{\epsilon}_I & i\tilde{\kappa}_I\\
\end{array} \right],\quad
\mathbf{J} = \left[
\begin{array}{cc}
1 & 0\\
0 & -1\\
\end{array} \right].
\end{eqnarray}

To proceed further, we note that the compatibility conditions required for Eqs.~(\ref{eq:zero-order})-(\ref{eq:second-order}) to be solvable, known also as Fredholm alternatives \cite{rpm1,hasbook1}, are $\mathbf{L}^{\pm}\mathbf{W}^{\pm}\mathbf{f}^{\pm}_{0,1,2} =0$,
where $\mathbf{L}^{\pm}=[1, \mp i\tilde{Z}_\ell]$ is a left eigenvector of $\mathbf{W}^{\pm}$,
such that $\mathbf{L}^{\pm} \mathbf{W}^{\pm}=[0,0]$, with $\tilde{Z}_\ell=\sqrt{\tilde{\mu}_R/\tilde{\epsilon}_R}$ being the
linear impedance, when dissipation is small enough to be ignored. Note that this impedance is independent of the chirality parameter $\tilde{\kappa}$ \cite{bookchiral2}.

The zeroth-order Eq.~(\ref{eq:zero-order}) provides the following three results. First, the solution $\mathbf{f}^{\pm}_0$  has the form:
\begin{eqnarray}
\label{eq:x0-result} \mathbf{f}^{\pm}_0=\mathbf{R}^{\pm}
\phi^{\pm}(Z,T^{\pm}),
\end{eqnarray}
where the scalar function $\phi^{\pm}(Z,T^{\pm})$ has to be determined
and $\mathbf{R}^{\pm}=[1, \mp i \tilde{Z}_\ell^{-1}]^{\rm T}$ is a right eigenvector of $\mathbf{W}^{\pm}$, i.e.,
$\mathbf{W}^{\pm}\mathbf{R}^{\pm}=[0,0]^{\rm T}$. Second, by using the compatibility condition
$\mathbf{L}^{\pm}\mathbf{W}^{\pm}\mathbf{f}^{\pm}_0 =0$ and Eq.~(\ref{eq:x0-result}), we obtain
$\mathbf{L}^{\pm}\mathbf{W}^{\pm}\mathbf{R}^{\pm}=0$, which is
actually equivalent to Eq.~(\ref{dr}). Third, the respective zeroth-order Beltrami components of the electric and magnetic fields
are proportional to each other, namely, $q^{\pm}_0 = \pm i \tilde{Z}_\ell p^{\pm}_0$ \cite{bookchiral2}.

Next, at $\mathcal{O}(\varepsilon^1)$, the compatibility conditions for Eq.~(\ref{eq:first-order}) result in
$\mathbf{L}^{\pm}(\mathbf{W}^{\pm})'\mathbf{R}^{\pm}=0$; equivalently,
\begin{eqnarray}
\label{eq:group-veloc}
2(ck_{c}^{\pm}\mp\omega_c\tilde{\kappa}_R)(c\tilde{k}^{\pm}\mp\omega\tilde{\kappa}_R)'\big|_{\omega=\omega_c}
=(\omega^2\tilde{\epsilon}_{R}\tilde{\mu}_{R})' \big|_{\omega=\omega_c}.
\end{eqnarray}
The foregoing equations yield the group speeds $v^{\pm}_g  \equiv
1/k_c^{\pm(1)}$,
which can also be derived from Eq.~(\ref{dr}). Furthermore,
Eqs.~(\ref{eq:first-order}) and (\ref{eq:x0-result}) together
indicate that $\mathbf{f}^{\pm}_1$ has the form:
\begin{eqnarray}
\label{eq:x1-result}
\mathbf{f}^{\pm}_1=i(\mathbf{R}^{\pm})'\partial_{T^{\pm}}\phi^{\pm}(Z,T^{\pm})+\mathbf{R}^{\pm}\psi^{\pm}(Z,T^{\pm}),
\end{eqnarray}
where the scalar function $\psi^{\pm}(Z,T^{\pm})$ can be determined at a higher-order approximation; see, e.g., Ref.~\cite{nlmms2c}.

Next, at order $\mathcal{O}(\varepsilon^2)$, the compatibility
conditions for Eq.~(\ref{eq:second-order}), combined with
Eqs.~(\ref{eq:x0-result}) and (\ref{eq:x1-result}), yield the
coupled NLS equations
\begin{eqnarray}
&i\partial_Z\phi^{\pm} -\frac{1}{2}k_c^{\pm(2)}\,
\partial^2_{T^{\pm}}\phi^{\pm}+\gamma\left(|\phi^+|^2+|\phi^-|^2\right)\phi^{\pm} =-i\tilde{\Gamma}^{\pm}\phi^{\pm},
\label{eq:NLS}
\end{eqnarray}
where $k_c^{\pm(2)}$ can be obtained from
Eq.~(\ref{eq:group-veloc}). In this equation, the nonlinearity
coefficient
\begin{eqnarray}
\label{eq:gamma} \gamma=\frac{\omega_c}{2}[\epsilon_0 (\alpha_1
\tilde{Z}_\ell+ \beta_1 \tilde{Z}_\ell^{-1}) +\mu_0 (\alpha_2
\tilde{Z}_\ell +\beta_2 \tilde{Z}_\ell^{-3})]
\end{eqnarray}
depends linearly on the scalar Kerr coefficients $\alpha_{1,2}$
and $\beta_{1,2}$, and the loss coefficients
$\tilde{\Gamma}^{\pm}$ are given as follows:
\begin{eqnarray}
\label{eq:Gamma}
\tilde{\Gamma}^{\pm}=\omega_c\left(
\frac{\tilde{\epsilon}_I\tilde{\mu}_R+\tilde{\epsilon}_R\tilde{\mu}_I}{2\sqrt{\tilde{\epsilon}_R\tilde{\mu}_R}}
\pm\frac{\tilde{\kappa}_I}{c}\right).
\end{eqnarray}

\subsection{Relation with the Manakov system}
Knowledge of $\phi^{\pm}$ obtained from solving Eqs. (\ref{eq:NLS}) immediately yields $q^{\pm}_0 = \phi^{\pm}$
and $p^{\pm}_0 = \mp i \tilde{Z}_\ell^{-1} \phi^{\pm}$, by virtue of
Eq.~(\ref{eq:x0-result}), similarly to the case of a linear material.

Let us now analyze Eqs.~(\ref{eq:NLS}) in more detail.
First, measuring length $z$, retarded time $T=t-z/v_g^{+}$, and the intensities
$|\phi^{\pm}|^2$ in units of a characteristic length $L^+=t_0^2 /|k^{+(2)}_c|$,
a characteristic time $t_0$, and $L^+/|\gamma|$, respectively, we reduce
Eqs.~(\ref{eq:NLS}) to the following dimensionless form:
\begin{eqnarray}
&i\partial_{Z}\phi^{+} -\frac{s}{2}
\partial_{T}^2 \phi^{+} +\sigma \left(|\phi^+|^2+|\phi^-|^2\right)\phi^{+}
=-i\Gamma^{+}\phi^{+},
\label{eq:NLS-scaled-1} \\
&i\left(\partial_{Z}\phi^{-}-\delta\,
\partial_{T}\phi^{-}\right) -\frac{d}{2}
\partial_{T}^2 \phi^{-} +\sigma \left(|\phi^+|^2+|\phi^-|^2\right)\phi^{-}=-i\Gamma^{-}\phi^{-}.
\label{eq:NLS-scaled-2}
\end{eqnarray}
The parameters involved in the foregoing equations are defined as
follows:
\begin{eqnarray}
\delta&=t_0\frac{ k^{+(1)}_c - k^{-(1)}_c}{|k^{+(2)}_c|},
\quad
d=\frac{k^{-(2)}_c}{|k^{+(2)}_c|}, \quad
s= \frac{k^{+(2)}_c}{|k^{+(2)}_c|},
\nonumber \\
\sigma&= \gamma/|\gamma|,
\quad
\Gamma^{\pm}=L^+\tilde{\Gamma}^{\pm}.
\label{param}
\end{eqnarray}
Since all nonlinearity coefficients in the $\pm$ components of
Eq.~(\ref{eq:NLS}) are equal to $\gamma$, the latter parameter was
absorbed by our normalization; this way, the nonlinearity
coefficients in the dimensionless Eqs.~(\ref{eq:NLS-scaled-1}) are
(\ref{eq:NLS-scaled-2}) are equal to the sign of $\gamma$.
Furthermore, as  $s={\rm sign}\left[k^{+(2)}_c\right]$, the
foregoing NLS system is characterized  by (i) the parameter
$\delta$, which represents the mismatch between the group speeds
of the two Beltrami components; (ii) the normalized dispersion
parameter $d$ for the field $\phi^{-}$; and (iii)
 the linear loss parameters $\Gamma^{\pm}$.

In the most general case ($\delta \ne 0$, $d \ne \pm 1$ and $\Gamma^{\pm} \ne 0$), the system of Eqs.~(\ref{eq:NLS-scaled-1})
and (\ref{eq:NLS-scaled-2}) is not integrable. The same is true even if $\Gamma^{+} =\Gamma^{-}=0$; however,
for certain values of the linear constitutive parameters and the nonlinearity coefficient $\gamma$, one can find, either analytically or numerically, various types of vector solitons (such as bright-bright, dark-dark, or dark-bright ones),
as can be gathered from Refs.~\cite{kiag,kivpr2}.

Provided $\delta=0$, $d=s$ and $\Gamma^{\pm}=0$, Eqs.~(\ref{eq:NLS-scaled-1}) and(\ref{eq:NLS-scaled-2}) reduce to the  Manakov system \cite{manakov}.
In our case, the latter takes the generic form of a vector NLS equation, which can be expressed as
\begin{eqnarray}
i \partial_Z {\bf u} - \frac{s}{2}\partial_{T}^2 {\bf u} + \sigma |{\bf u}|^2 {\bf u} = \bf{0},
\label{man}
\end{eqnarray}
where ${\bf u}(Z,T)=\left[\phi^{+}(Z,T), \phi^{-}(Z,T)\right]^T$. The Manakov system is known to be completely integrable \cite{intman1,intman2,intman3}; in fact, it can be integrated by extending the inverse scattering transform method that has been used to integrate the scalar NLS equation \cite{zsbd1,zsbd2}. The Manakov system also admits vector $N$-soliton solutions \cite{ralak,jy1,jy2,kivtur,shepkiv,park}.

In the following section, we show that it is possible to find physically relevant conditions allowing us to approximate the general system of
the NLS Eqs.~(\ref{eq:NLS-scaled-1}) and (\ref{eq:NLS-scaled-2}) to the Manakov system (\ref{man}) and investigate conditions for the existence of various types of vector solitons supported by a nonlinear isotropic chiral material.

\section{Vector solitons}\label{sec3}

We considered a certain type of isotropic chiral material that
conforms to the
 single-resonance Lorentz models for the linear parts of permittivity and  permeability, and to the
 Condon model for the chirality parameter (see, e.g., Ref.~\cite{lakhtakia-josaa} and references therein) as follows:
\begin{eqnarray}
\tilde{\epsilon}_\ell (\omega)&=&\epsilon_0 \left( 1-\frac{\omega_{p}^2}{\omega^2 -
\omega_{\epsilon}^2 + 2i\delta_{\epsilon} \omega} \right),
\label{epsL} \\
\tilde{\mu}_\ell (\omega)&=&\mu_0 \left( 1-\frac{\omega_{m}^2}{\omega^2 -\omega_{\mu}^2 + 2i\delta_{\mu} \omega} \right),
\label{muL} \\
\tilde{\kappa} (\omega)&=& \frac{a c Z_0^{-1} \omega}{\omega^2 -\omega_{\kappa}^2 + 2i\delta_{\kappa} \omega}.
\label{kL}
\end{eqnarray}
Here, $Z_0 = \sqrt{\mu_0/\epsilon_0}$ is the intrinsic  impedance
of free space; $\{\omega_{\epsilon}, \delta_{\epsilon}\}$,
$\{\omega_{\mu}, \delta_{\mu}\}$, and $\{\omega_{\kappa},
\delta_{\kappa}\}$ are the resonance angular frequencies and
linewidths of $\tilde{\epsilon}_\ell$, $\tilde{\mu}_\ell$, and
$\tilde{\kappa}$, respectively; while $\omega_{p}$ and
$\omega_{m}$ characterize the oscillator strengths of respective
transitions in $\tilde{\epsilon}_\ell$ and $\tilde{\mu}_\ell$.
Finally, $a$ is the rotatory strength of the resonance in the
Condon model, which measures the degree of chirality.
We note that the Lorentz model is a widely used fundamental model,
describing the dispersive nature of permittivity and permeability,
also accounting for the presence of a single resonance frequency.
On the other hand, the Condon model is generally used to represent
the dispersive nature of chirality; the Condon model in the form
of Eq.~(\ref{kL}) is such that the real (imaginary) part of
$\tilde{\kappa}(\omega)$ is an odd (even) function of frequency
\cite{Emeis1967,GL1995},  and obeys causality restrictions.

Next, we used Eqs.~(\ref{epsL}), (\ref{muL}) and (\ref{kL}), as
well as the dispersion relations (\ref{dr}), to determine the
refractive indices $\tilde{n}^{\pm}$ [cf. Eq.~(\ref{cri})] and the
parameters involved in the NLS equations [cf. Eq.~(\ref{param})].
All numerical results presented were computed for the following
values of parameters: $\omega_p = 0.9 \omega_{\epsilon}$,
$\omega_m = 0.8 \omega_{\epsilon}$, $\omega_{\mu} = 0.8
\omega_{\epsilon}$, $\omega_{\kappa} = 0.6 \omega_{\epsilon}$,
$\delta_{\epsilon} =\delta_{\mu} = \delta_{\kappa} = 10^{-3}
\omega_{\epsilon}$, and $a=0.5 \omega_{\epsilon}/c$. Note that all
resonance angular frequencies and linewidths are normalized to the
resonance angular frequency $\omega_{\epsilon}$ of
$\tilde{\epsilon}_\ell$, which was taken to be the largest
resonance angular frequency in the constitutive description of the
chiral material. Other parameter values led to qualitatively
similar results.

\begin{figure}
\hspace{2.3cm}
\includegraphics[scale=0.33]{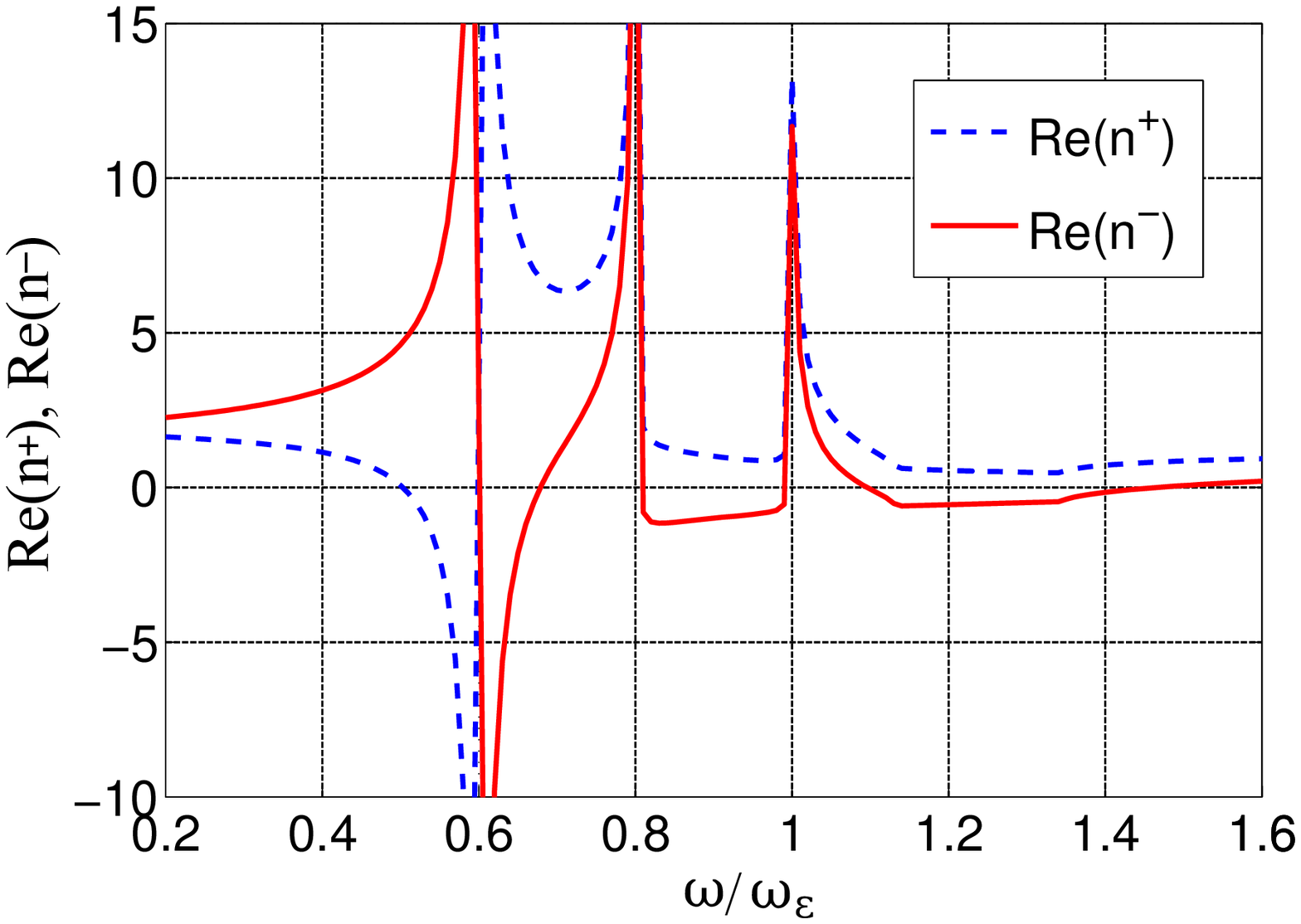}
\includegraphics[scale=0.4]{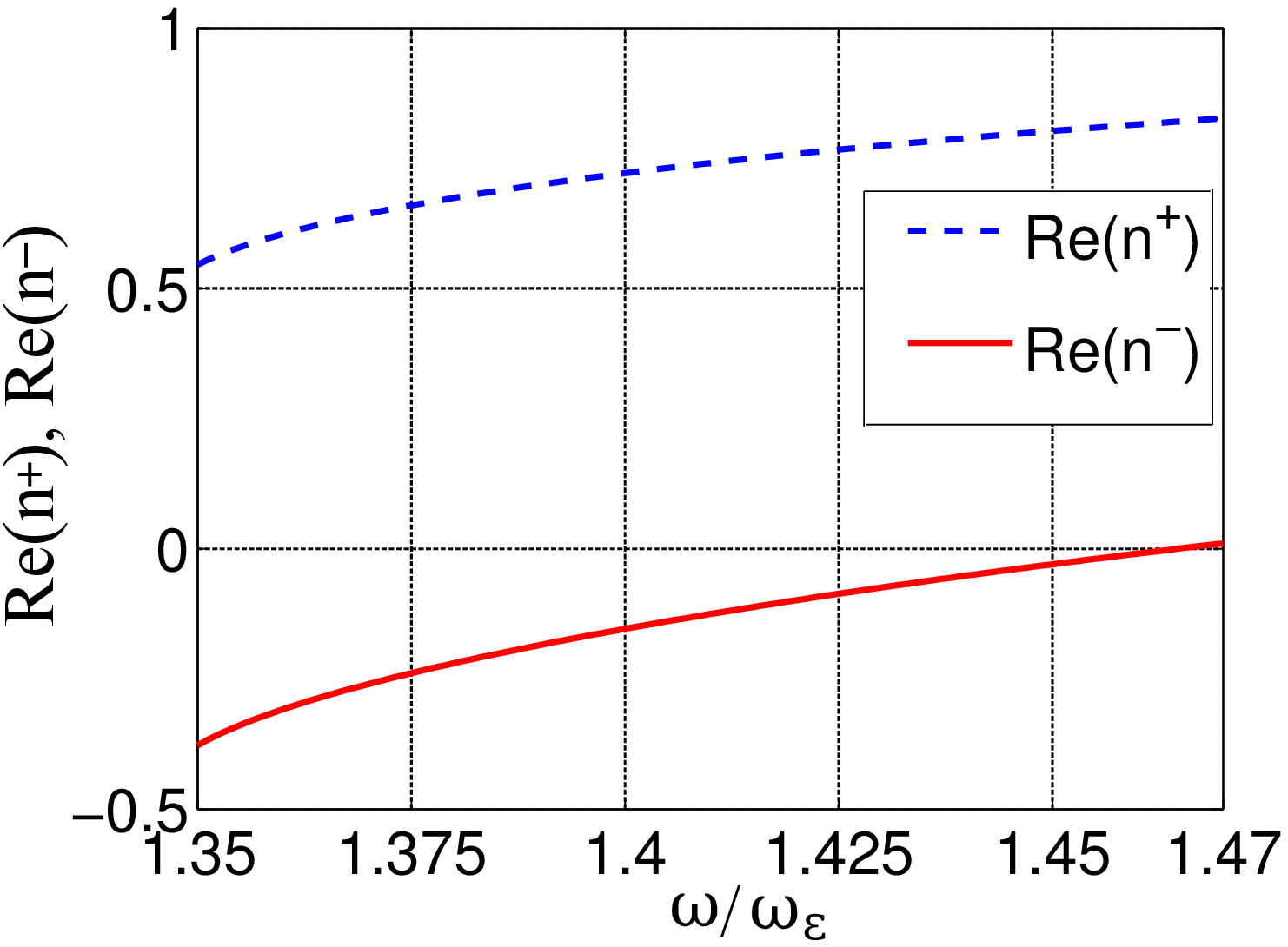}
\caption{(Color online) Left panel: ${\rm Re}(\tilde{n}^{-})$
(solid red line) and ${\rm Re}(\tilde{n}^{+})$ (dashed blue line)
in the spectral regime $(0.2 \omega_{\epsilon}, 1.6
\omega_{\epsilon})$. The constitutive parameters used for these
data are presented at the beginning of Sec.~\ref{sec3}.  Three distinct spectral regimes exist
wherein ${\rm
Re}(\tilde{n}^{-})<0$ and ${\rm Re}(\tilde{n}^{+})>0$.
Right panel: Magnification of left panel in the spectral subregime $(1.35
\omega_{\epsilon}, 1.47 \omega_{\epsilon})$, wherein not only are ${\rm
Re}(\tilde{n}^{-})<0$ and ${\rm Re}(\tilde{n}^{+})>0$,
but the NLS
Eqs.~(\ref{eq:NLS-scaled-1}) and (\ref{eq:NLS-scaled-2}) can also be
approximated by the Manakov system (\ref{man}). }
\label{fig1}
\end{figure}

For the chosen parameters with $a>0$,  spectral regimes
 exist where
the refractive indices are $\tilde{n}^{-} <0$ and $\tilde{n}^{+}
>0$ if $a \gtrsim a_{c1} = 5 \times 10^{-3} \omega_{\epsilon}/c$.
In the top panel of Fig.~\ref{fig1} we show an example of the
frequency dependences of $\tilde{n}^{-}$ and $\tilde{n}^{+}$ for
$a = 0.5 \omega_{\epsilon}/c$. In fact, if the chirality parameter
is sufficiently large, i.e., $a \gtrsim a_{c2} = 0.1
\omega_{\epsilon}/c$, then there exists a certain spectral subregime
located to
the right of the largest resonance angular frequency
$\omega_{\epsilon}$, such that $\tilde{n}^{-} <0$, $\tilde{n}^{+} >0$, and the NLS
Eqs.~(\ref{eq:NLS-scaled-1}) and (\ref{eq:NLS-scaled-2}) can be
approximated by the Manakov system (\ref{man}). Hence, this
spectral regime is referred to as the \textit{Manakov regime} in
the remainder of this paper. In the particular case of $a=0.5
\omega_{\epsilon}/c$, the Manakov regime is
$(1.35\omega_{\epsilon}, 1.47\omega_{\epsilon})$; the dependencies
of $\tilde{n}^{-}$ and $\tilde{n}^{+}$ on $\omega$ in this regime
are illustrated in the bottom panel of Fig.~\ref{fig1}. Here we
should note that increase of $a$ above the characteristic value
$a_{c2}$ results in the increase of the bandwidth of the Manakov
regime; for example, when $a=1$, the Manakov regime is
$(1.35\omega_{\epsilon}, 1.73\omega_{\epsilon})$.

If the sign of $a$ is changed, then a role-reversal occurs in that
$\tilde{n}^{+} <0$ and $\tilde{n}^{-} >0$ in the Manakov regime.
Although we have maintained $a>0$ in the remainder of this
section, the case of $a <0$ can be treated analogously.

\begin{figure}
\hspace{2.3cm}
\includegraphics[scale=0.4]{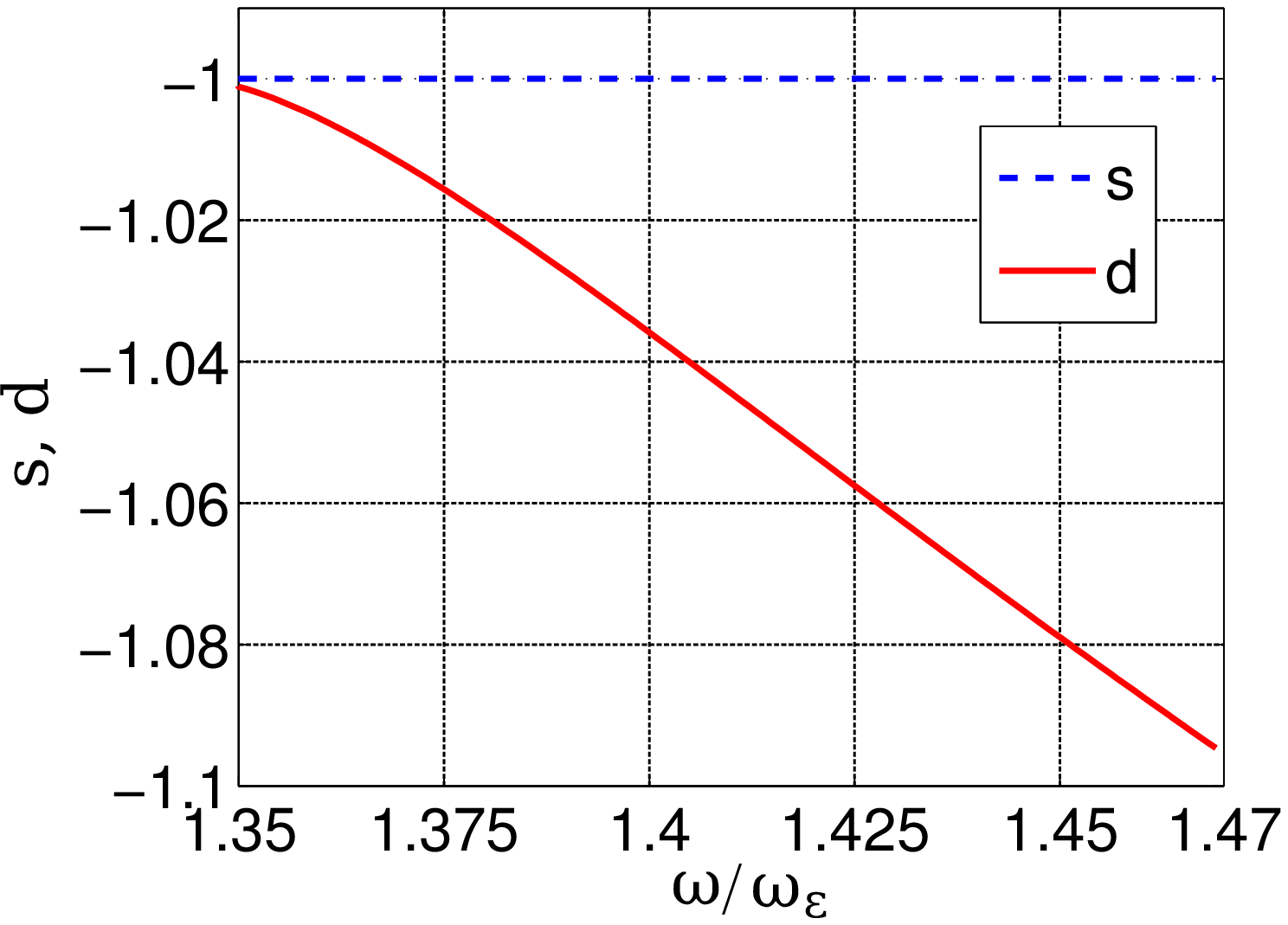}
\includegraphics[scale=0.4]{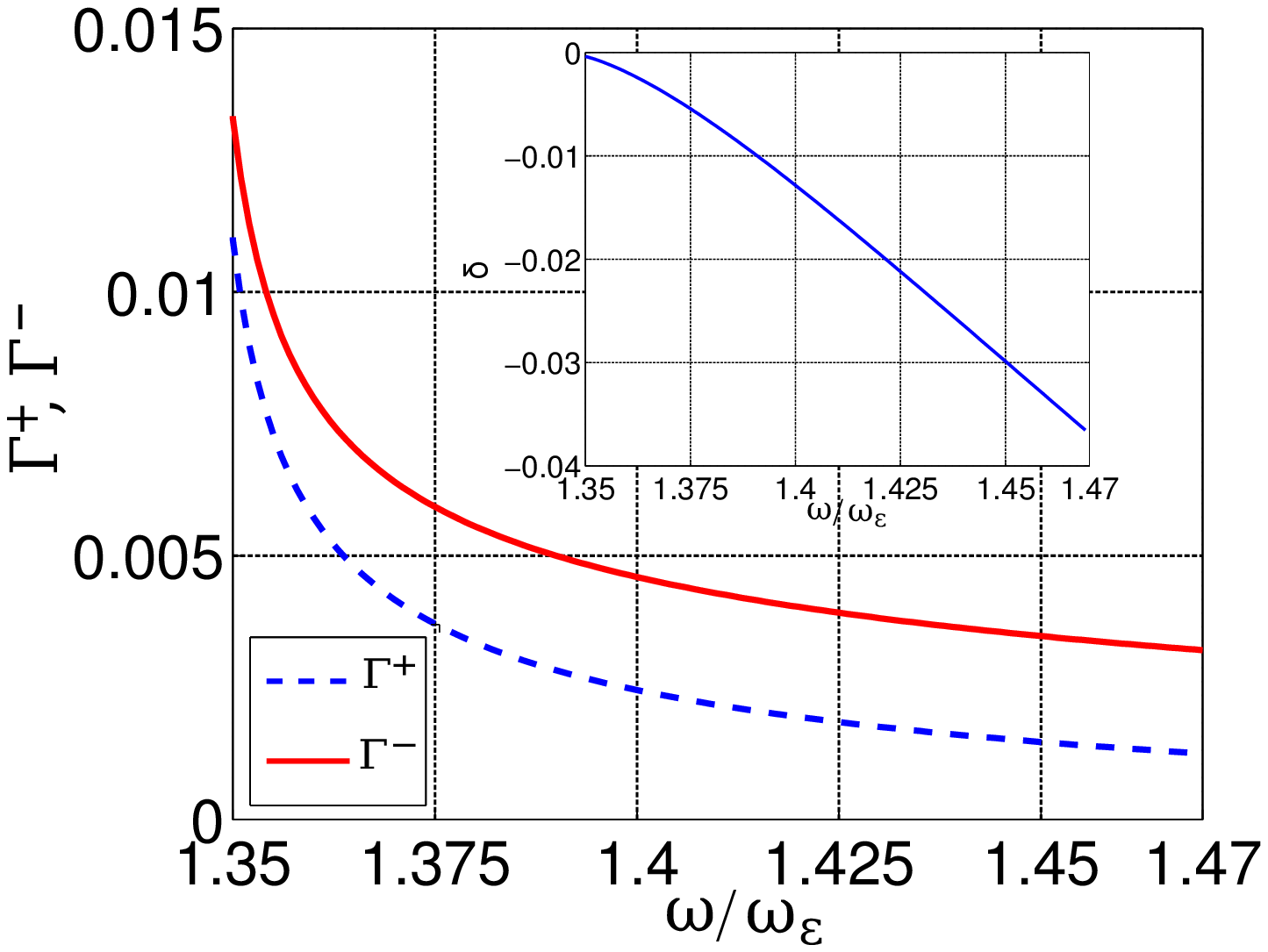}
\caption{(Color online) Left panel: Normalized dispersion
coefficients $s$ (dashed blue line) and $d$ (solid red line) of
the NLS Eqs.~(\ref{eq:NLS-scaled-1})-(\ref{eq:NLS-scaled-2}),
defined in Eqs.~(\ref{param}), in the Manakov regime $(1.35
\omega_{\epsilon}, 1.47 \omega_{\epsilon})$. Right panel:
Normalized loss coefficients, $\Gamma^{+}$ (dashed blue line) and
$\Gamma^{-}$ (solid red line) in the Manakov regime. Inset shows
the group-speed-mismatch coefficient $\delta$. } \label{fig2}
\end{figure}

In the Manakov regime $d \approx s$, while $\delta \approx 0$ and
$\Gamma^{\pm} \approx 0$. This is confirmed by the data presented
in Fig.~\ref{fig2}. Whereas $d\approx -1$, the maximum deviation
between $s$ and $d$ (for $\omega = 1.47 \omega_{\epsilon}$) is
than $10\%$. Furthermore, $\Gamma^{\pm}$ are of order of $10^{-3}$
while $\delta$ is of order of $10^{-2}$. Thus, taking into account
that the other coefficients in Eqs.~(\ref{eq:NLS-scaled-1}) and
(\ref{eq:NLS-scaled-2}) are equal to $\pm 1$, we can safely
conclude that, in a first approximation, our system can be well
approximated by the Manakov system (\ref{man}).

Here it should be noted that, following the foregoing arguments,
the (small) effects of the group-speed mismatch between the two
Beltrami components and the linear losses can be studied
analytically by means of the perturbation theory for solitons;
see, e.g., Ref.~\cite{RMP}. However, such a study is beyond the
scope of this work.
In any case, some estimations concerning, e.g., the role of losses
can already be made, based on the linear, dispersionless part of
the NLS Eqs.~(\ref{eq:NLS-scaled-1})-(\ref{eq:NLS-scaled-2}): one
expects that the Beltrami fields $\phi^{\pm}$ decay as
$\exp(-\Gamma^{\pm}Z)$. To further elaborate on this estimation,
however, one would employ real world parameter values, which are
not currently available (since experiments on isotropic nonlinear
chiral metamaterials considered herein have not been performed so
far).

The reduction of Eqs.~(\ref{eq:NLS-scaled-1}) and
(\ref{eq:NLS-scaled-2}) to the Manakov system allows us to predict
different types of exact vector solitons that can propagate in an
isotropic chiral NRRI material. In particular, taking into regard
that $s=d=-1$, there are two different cases, depending on the
sign $\sigma$ of the nonlinearity coefficient $\gamma$:
\begin{itemize}
\item[(a)] $\sigma=+1$, i.e., the effective Kerr nonlinearity in the Manakov system is {\it self-focusing}, and
\item[(b)] $\sigma=-1$, i.e., the effective Kerr nonlinearity is {\it self-defocusing}.
\end{itemize}

\subsection{Self-focusing nonlinearity in the Manakov regime}\label{sfnM}

When $s=d=-1$, $\sigma = +1$, and the boundary conditions
associated with Eq.~(\ref{man}) are $\phi^{\pm} \rightarrow 0$ as
$T\rightarrow 0$, there exist exact bright-bright soliton
solutions, i.e., both Beltrami components take the form of bright
solitons. Following Refs.~\cite{jy1,jy2},
we may find such solutions by using the traveling-wave ansatz
\begin{eqnarray}
\phi^{+}(Z,T) &=& u_1(\xi)\exp\left[i(1+C^2/2)Z +i CT \right],
\label{BB1} \\
\phi^{-}(Z,T) &=& u_2(\xi)\exp\left[i(q^2+C^2/2)Z +i CT \right],
\label{BB2}
\end{eqnarray}
where $\xi = \sqrt{2}(T-CZ)$, the parameter $C$ sets the (common)
speed of both components of the traveling wave in the
$(Z,T)$-plane, $q$ is an arbitrary parameter connected with the
wave number of the right-handed Beltrami component, and $u_{1,2}$
are unknown real functions.

Introducing the ansatz comprising Eqs.~(\ref{BB1}) and (\ref{BB2}) into the Manakov
system (\ref{man}), we obtain the following system of ordinary
differential equations (ODEs) for $u_1$ and $u_2$:
\begin{eqnarray}
&&\frac{d^2u_1}{d\xi^2} -u_1 + (u_1^2 +u_2^2)u_1=0,
\label{em1} \\
&&\frac{d^2u_2}{d\xi^2} -q^2 u_2 + (u_1^2 +u_2^2)u_2=0.
\label{em2}
\end{eqnarray}
The simplest exact solution to this system has either $u_1\equiv0$
or $u_2\equiv0$, i.e., either
\begin{eqnarray}
u_1 &=& 0, \quad u_2=\sqrt{2}q {\rm sech}(q \xi)
\label{sbb1-1}
\end{eqnarray}
or
\begin{eqnarray}
u_1 &=& \sqrt{2} {\rm sech}(\xi), \quad u_2=0.
\label{sbb1-2}
\end{eqnarray}

This solution can be generalized to the case wherein both $u_1$
and $u_2$ are nontrivial \cite{jy1,jy2}. Particularly, if $q=1$, a
simple one-parameter family of such solutions is composed of
symmetric, single-humped bright solitons of the following form:
\begin{eqnarray}
u_1= \sqrt{2} \cos\theta {\rm sech}(\xi), \quad u_2= \sqrt{2} \sin\theta {\rm sech}(\xi),
\label{bb}
\end{eqnarray}
where $\theta$ is an arbitrary parameter. On the other hand, if
$0<q<1$, the soliton is
\begin{eqnarray}
u_1&=& \frac{\sqrt{2(1-q^2)}\cosh(q\xi)}{\cosh(\Delta \xi)\cosh(q\xi)-q\sinh(\Delta \xi)\sinh(q\xi)},
\label{bbb1} \\
u_2&=& \frac{-q\sqrt{2(1-q^2)}\sinh(\Delta \xi)}{\cosh(\Delta \xi)\cosh(q\xi)-q\sinh(\Delta \xi)\sinh(q\xi)},
\label{bbb2}
\end{eqnarray}
where $\Delta \xi \equiv \xi-\xi_0$ and $\xi_0$ is an arbitrary
constant. The solutions (\ref{bbb1}) and (\ref{bbb2}) are generally
asymmetric; however. $u_1$ is symmetric but $u_2$ is antisymmetric
when $\xi_0=0$. Other solution branches, such as soliton bound
states and wave- and daughter-wave solutions (for $u_1 \ll u_2$ or
$u_2 \ll u_1$), have also been found; see, e.g.,
Refs.~\cite{jy1,jy2}, as well as the book
\cite{kiag} and references therein.

\subsection{Self-defocusing nonlinearity in the Manakov regime}\label{sdnM}

Let us now change the sign of $\sigma$ to
 $\sigma=-1$ and keep   $s=d=-1$. Now, there exist both dark-dark solitons
 (where both the Beltrami components take the form of dark solitons)
 and dark-bright ones (where one Beltrami component is a dark soliton and the other is a bright soliton).

Dark-dark soliton solutions of the Manakov system (\ref{man}), with boundary conditions $|\phi^{\pm}| \rightarrow \mu_{\pm}$ as $T\rightarrow 0$, where $\mu_{\pm}$ are the amplitudes of the continuous-wave (cw) backgrounds carrying the dark solitons, can be expressed in the following general form \cite{shepkiv,kivtur}:
\begin{eqnarray}
&&\phi^{\pm}(Z,T) = \mu_{\pm}\{ \cos \varphi_{\pm} \tanh[\nu(T-V Z)] + i \sin \varphi_{\pm} \} \nonumber \\
&&\times \exp\left[i \left(\frac{\eta_{\pm}^2}{2} + |\mu_{+}|^2 + |\mu_{-}|^2 \right)Z + i \eta_{\pm}T \right],
\label{dd}
\end{eqnarray}
where $\nu$ and $V$ are the (common) inverse width and speed of
dark solitons, while $\eta_{\pm}$ are constants connected to the
background angular frequency $\omega_c$ and wave numbers $k^\pm_c$.
The phase angles $\varphi_{\pm} = \tan^{-1}[(\eta_{\pm}-V)/\nu]$
are connected with the rest of the soliton parameters according to
the relations $\nu^2 = \mu_{+}^2 \cos^2\varphi_{+}+\mu_{-}^2
\cos^2\varphi_{-}$, while the soliton intensities are as follows:
\begin{eqnarray}
|\phi^{\pm}|^2 = \mu_{\pm}^2-\nu^2 {\rm sech}^2[\nu(T-V Z)].
\label{idd}
\end{eqnarray}

Finally, for the same case ($\sigma=-1$), we present the exact
analytical dark-bright soliton solutions of the Manakov system
(\ref{man}) \cite{ralak,shepkiv}:
\begin{eqnarray}
\phi^{+}(Z,T)&=&\phi_0\{\cos\varphi \tanh[D(T-C_s Z)] + i\cos\varphi\}
\nonumber \\
&\times&\exp[i\Omega T +i (\Omega^2 + \phi_0^2)Z],
\label{db1} \\
\phi^{-}(Z,T)&=&\eta\, {\rm sech}[D(T-C_s Z)]
\nonumber \\
&\times&\exp\{iC_s T +i [(D^2 - C_s^2)/2 - \phi_0^2]Z \},
\label{db2}
\label{idd}
\end{eqnarray}
where $\phi^{+}$ and $\phi^{-}$ correspond, respectively, to the dark and bright soliton components;  $\phi_0$ is the amplitude of the cw background carrying the dark soliton; $\varphi=\tan^{-1}[(\Omega-C_s)/D]$ is the dark soliton's phase angle; $D$ and $C_s$ denote the (common) inverse width and speed of both components; $\Omega$ sets the frequency of the dark-soliton component; and $\eta$ is the bright-soliton amplitude, with $\eta^2 = \phi_0^2 \cos^2\varphi - D^2$. In the limiting case of $\eta\rightarrow 0$, the bright soliton component vanishes and the inverse width of the dark soliton becomes $D=\phi_0 \cos\varphi$; this way, Eq.~(\ref{db1}) describes the  single dark soliton solution of the defocusing NLS equation \cite{kivpr1,kivpr2}. It is important to stress here that the dark-bright soliton of Eqs.~(\ref{db1}) and (\ref{db2}) is a generic example of a  \textit{symbiotic soliton}: the bright-soliton component exists only due to the coupling with the dark soliton and would not be supported otherwise, due to the fact that the NLS equation for $\phi^{-}$ features a {\it self-defocusing} nonlinearity.

Finally, we note in passing that other soliton solutions of the Manakov Eqs.~(\ref{man}) can be systematically constructed by employing the symmetries of the system (such as the SU(2) rotational symmetry) \cite{park}.

\section{Conclusions}\label{sec4}

In conclusion, we have studied the propagation of electromagnetic
pulses in isotropic chiral NRRI materials with Kerr nonlinearity.
We have employed the reductive perturbation theory to derive a
system of two coupled nonlinear Schr\"{o}dinger (NLS) equations
for the envelopes of the two Beltrami components of the EM field.
Assuming a single-resonance double Lorentz model for the effective
permittivity and permeability of the medium, as well as a Condon
model for the chirality parameter, we have shown that if the
characteristic chirality parameter is sufficiently large, then the
material exhibits a negative-real refractive index for the
right/left-handed Beltrami component in certain spectral regimes
whereas the left/right-handed Beltrami component does not.

Furthermore, we have shown that in a certain subregime inside
such a spectral regime, the system of the NLS equations
 (which is non-integrable in general) may be approximated well by the completely integrable Manakov system.
 This way, we have predicted various types of vector solitons, such as bright-bright, dark-dark or dark-bright ones,
 that can be formed in the chosen type of chiral material. Let us emphasize that, in all cases, the presented vector
 solitons are composed by a Beltrami component that features negative-real refractive index
 and another Beltrami component that features a positive-real refractive index.

The propagation of the predicted vector soliton solutions needs to
be examined by direct numerical simulations in the framework of
the Maxwell equations. Furthermore, a relevant future challenge is
the analysis of pulse propagation in the bianisotropic
\cite{ml2004, ml2008} counterparts of the isotropic materials we
have discussed here. For that purpose, our work can be emulated in
three steps as follows. In the first step, one has to obtain
explicit plane wave solutions and associated dispersion relations,
similar to the ones in Eqs.~(\ref{dr}). In the secon step, one has
to consider propagation of wavepackets, similar to Eqs.
(\ref{eq:Efield-decomp}) and (\ref{eq:Hfield-decomp}), along a
fixed direction (identified by the $z$-axis in our work) of the
linearized material~---~see, e.g., Ref.~\cite{ml2008}. The third
and final step is to apply the methodology of Section~2.3 to
derive a vector NLS-type model (and, in the best-case scenario, a
Manakov system) describing nonlinear pulse propagation in the
nonlinear material.
Also, another interesting possibility would be to apply the
derived results in the case of pulse propagation inside a layered
chiral \cite{LVV1989,LVV1990} or a layered bianisotropic material
\cite{Tsalamengas,KKNE}.
For this purpose, one should first find expressions for the EM
field components inside each homogeneous layer (relying, e.g., on
plane waves for the linear layers, and nonlinear periodic waves or
solitons for the nonlinear layers) and then impose the boundary
conditions at the interfaces. Finally, our work may be exploited
for the analysis of pulse propagation inside composite materials
comprising linear chiral and nonlinear dielectric materials
\cite{Lak-Weig}. Such studies are currently in process and
relevant results will be reported in the future.

\section*{Acknowledgments.} A.L. thanks the Charles Godfrey Binder Endowment at Penn State for partial support of his research.
The work of D.J.F. was partially supported by the Special Account for Research Grants of the  University of Athens.

\end{document}